# Scanning tunneling microscopy and Raman evidences of silicene nanosheets intercalated into graphite surface at room temperature

I.Kupchak[1], F.Fabbri[2], M.De Crescenzi[3], M. Scarselli[3], M. Salvato[3], T.Delise[3], I.Berbezier[4], O.Pulci[3] and P. Castrucci[3,*]

[1] V. E. Lashkaryov Institute of Semiconductor Physics, National Academy of Sciences of Ukraine, Ukraine

[2] NEST, Scuola Normale Superiore, Istituto Nanoscienze-CNR, Pisa, 56127, Italy.

[3] Dipartimento di Fisica, Università di Roma "Tor Vergata", 00133 Roma, Italy.

[4] CNRS, Aix-Marseille Université, IM2NP, UMR 7334, Campus de St. Jérome, 13397 Marseille, France.

**Abstract:**

Highly oriented pyrolitic graphite (HOPG) is an inert substrate with a structural honeycomb lattice, well suited for the growth of two-dimensional silicene layer. It was reported that when Si atoms are deposited on HOPG surface at room temperature, they arrange in two configurations: silicene nanosheets and three dimensional clusters. In this work we demonstrate by means of scanning tunneling microscopy (STM) and Raman spectroscopy that a third configuration stabilizes in the form of Si 2D nanosheets intercalated below the first top layer of carbon atoms. The Raman spectra reveal a structure located at 538 cm$^{-1}$ which we ascribe to the presence of sp$^2$ Si hybridization. In addition, the silicon deposition induces several modifications in the D and G Raman modes of graphite, interpreted as an experimental evidence of the intercalation of the silicene nanosheets. The Si atom intercalation at room temperature takes place at the HOPG step edges and it detaches only the outermost graphene layer inducing a strong tensile strain mainly concentrated on the edges of the silicene nanosheets. Theoretical calculations of the structure and energetic viability of the silicene nanosheets, of the strain distribution on the outermost graphene layer and its influence on the Raman resonances support the experimental STM and Raman observations.

Two-dimensional (2D) materials commonly possess unique optical bandgap structures, extremely strong light–matter interactions, and large specific surface area. Graphene, hexagonal boron nitride (h-BN) and transition metal dichalcogenides (TMDs) have emerged as promising building blocks for novel nanoelectronics, providing a full range of material types, including large band gap insulators, semiconductors and semimetals.[1] Among these 2D materials, silicene is gaining increasing interest for the application in nanoelectronics and spintronics.[2]

Differently from graphene, that is one-atom thick carbon sheet with honeycomb $sp^2$ configuration, silicon generally takes tetravalence states like carbon in diamond structure although it belongs to the same group as carbon in the periodic table. In strong contrast with carbon, silicon tends to form $sp^3$ configuration with the surrounding silicon atoms in solid phase. In fact, silicon atoms deposited on a solid surface agglomerate, even at very low coverages, and form three-dimensionally diamond-like structure with $sp^3$ bonds.[3,4] In spite of these generally observed features, theoretical works have shown that there exists a metastable phase of single-layered silicon with graphene-like structure, called silicene, either in a flat or in a slightly puckered configuration.[5-8] The existence of this aggregation state has attracted much attention for the evident improvements that its use can imply in silicon based nano-electronic devices[9] compared to graphene. Recently some of us have reported evidence that patches of silicene can form on inert highly oriented pyrolitic graphite (HOPG).[10,11] In these works, scanning tunneling microscopy (STM) and *ab initio* molecular dynamics simulations revealed the growth of silicon nanosheets where the substrate/silicon interaction is minimized. STM measurements clearly display silicene nanosheets localized very close to small nanosized Si clusters leaving large part of HOPG free of additional nanostructures. High resolution STM images show both the atomically resolved unit cell and the presence of a small buckling in the silicene honeycomb structure. In this paper we report evidence that for the same amount of silicon, deposited at room temperature, a third possibility exists, i.e. Si atom intercalation under the first graphitic sheet of HOPG. STM measurements show that an exchange of Si atoms occurs, involving from one to many atoms up to the formation of nanometric sized modulations of the flat HOPG surface (hereafter called bubbles), thus decoupling the first HOPG layer from the substrate. Interestingly, these bubbles have a flat profile, monoatomically high and exhibit the same network and lattice periodicity of HOPG. All these observations support the formation of silicene nanosheets located underneath the first graphene layer of HOPG. Raman spectra reveal the presence of a feature located around 538 cm$^{-1}$, which we assign to these silicene bubbles.[9,12,13] In addition, the silicon deposition induces several modification to graphite D and G Raman modes. We interpret these modifications as a clear experimental evidence of the intercalation of the silicene nanosheets which induces tensile strain in the overneath graphene layer. Robust theoretical calculations on the structure and stability of the silicene nanosheets sandwiched among graphene and graphite layers have been done to support the experimental STM. Raman calculations performed for the carbon G band confirmed the G band feature huge shift induced by the tensile strain. This is one of the first evidence of Si intercalation at room temperature in graphite substrates.

A HOPG (from GE Advanced Ceramics, USA, 12 mm × 12 mm × 1 mm) sample was used as a substrate. A fresh surface of graphite was obtained by peeling the HOPG substrate with adhesive tape and transferring it into an ultra-high vacuum (UHV) chamber. Silicon atoms were evaporated from a wafer (Sil'tronix ST, ρ = 1−10 Ω•cm, n-doped) located at 200 mm from the substrate. The deposition was achieved under UHV conditions (base pressure low $10^{-10}$ Torr) and at a constant rate of 0.1 nm/min (0.04 ML/min) monitored by an Inficon quartz balance. Deposition was carried out keeping the substrate at room temperature (RT). STM imaging was performed using an Omicron-STM system with electrochemically etched tungsten tips. The STM was calibrated by acquiring atomically resolved images of the bare HOPG surface. All images were acquired in the constant current mode and were unfiltered apart from the rigid plane subtraction. Ex-situ Raman spectroscopy and mapping have been performed without capping the sample. The Raman analysis was carried out with a 532 nm excitation laser, a laser power density of 0.1 mW/μm$^2$ and an acquisition time of 1 s. The spectral resolution is 2 cm$^{-1}$. The Raman system is a Renishaw Invia Qotor equipped with a confocal optical microscope and a high resolution Andor CCD camera. From the fitting of the Raman features by using Voigt curves, it is possible to evaluate the peak position, full width at half maximum (FWHM) and its associated error (0.5 cm$^{-1}$).

Theoretical calculations on atomic structure of a silicon sheet under the graphite first layer have been performed within density functional theory (DFT) and generalized gradient approximation (GGA) including van der Waals corrections in form vdW-DF2 functional,[14] as implemented in the QUANTUM ESPRESSO software package.[15] We have used Perdew-Burke-Ernzerhof (PBE)[16] pseudopotentials in a projector augmented-wave form. An integration of the Brillouin zone was performed using a 3×3×1 Γ-centered grid of special points in k-space, generated by the Monkhorst-Pack scheme[17] and Methfessel-Paxton smearing[18] of 0.005 Ry. Convergence of the results was achieved with a 40 Ry cutoff in the wave functions and 160 Ry in the augmented charge density. We modeled the HOPG sample by a three-layer slab of a size 12×12 graphene unit cells, with a vacuum layer of 2.0 nm thickness in the z-direction to separate the periodic images of the slabs. The atomic layers were relaxed until the Hellmann-Feynman forces became less than 10-3 a.u., keeping the bottom layer fixed. Starting with a single Si atom introduced under the top graphite layer of C-slab, we consequently increased step by step the number of Si-atoms at the positions corresponding to free-standing silicene sites. For small amount of Si-atoms, the cluster tends to form fcc(111) plane during the geometry optimization. However, starting from 24 atoms, an internal part of Si-cluster preserves an initial, silicene-like structure, and such an atomic ordering is observed for the cluster with up to 37 Si-atoms.

In Fig. 1 (central panel), we report the STM image of HOPG surface exhibiting the atomic network typical of graphite surface with some regions brighter than others. This has been never observed for clean HOPG surface on extended nanometric terraces, while it appears soon after Si atom deposition. In the STM image it is possible to identify more brilliant areas involving one single atom, lines of atoms up to regions involving several atoms. By measuring the height of these brighter regions (see line profiles in top and bottom panels of Fig.1), we observe that the most brilliant spots have a vertical distance from the HOPG basal plane of about 0.15 nm. Moreover, the atomic distance still remains the same of the HOPG lattice (i.e. about 0.25 nm). It is worth noting that if the Si atom was on top of the C atom in the HOPG network as predicted by Aktürk et al.,[19] the line profile would be different from the one that we measured. In particular, we should have seen only one spot located at higher distance from the C atoms of the graphite surface instead of degrading heights of the surrounding atoms as shown in the reported line profile of Fig.1.

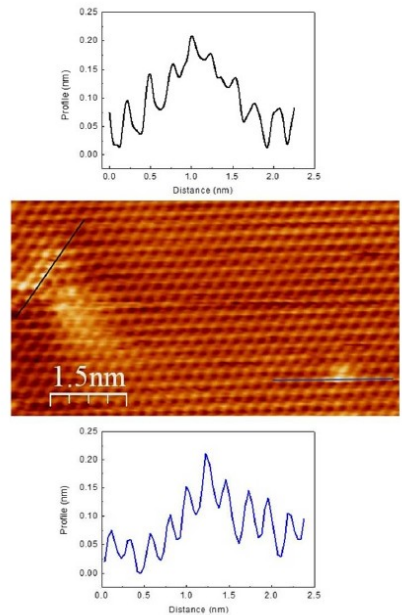

Figure 1: Central panel: STM image ($I_{tunn}$= 0.4 nA; $V_{bias}$= 0.3 V) of HOPG surface after the 1 ML Si deposition at room temperature. Upper and lower panels: height profiles of the brightest regions exhibiting a vertical distance from the HOPG basal plane of about 0.15 nm, while the planar distance between the spots remains the one typical of HOPG.

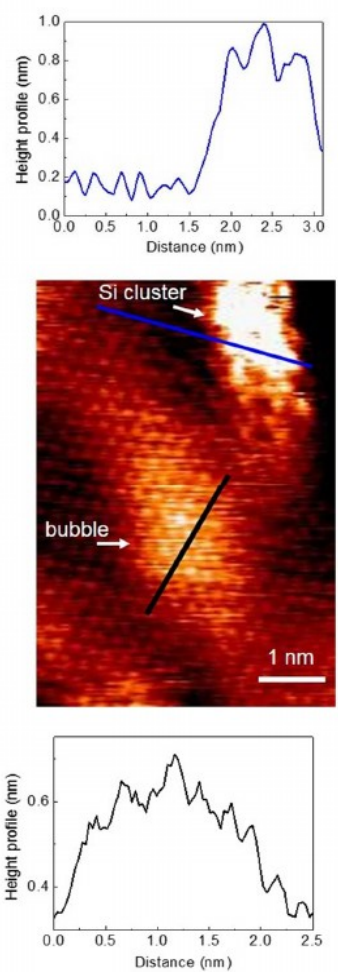

Figure 2: Central panel: STM image ($I_{tunn}$= 0.4 nA; $V_{bias}$= 0.3 V) of HOPG surface after the 1 ML Si deposition at room temperature, showing two different morphologies: a Si three-dimensional cluster and a two-dimensional bubble. The height of these features are very different as reported in the height profile shown in the top and lower panels.

In Fig. 2 we report an STM image displaying a Si cluster and a nanometer sized brilliant area. An STM image, exhibiting the coexistence of these two morphologies over a more extended area, is reported in Fig. S1 of the Supplementary Information (SI). A marked difference between the two regions is evidenced by the corresponding height profiles. The former (blue curve) shows a distance between the atoms much larger than that typical of HOPG and a distance from the basal plane of about 1 nm. The atoms constituting the latter area (black curve) have a height distance of a few tenths of nanometer and maintain the HOPG atomic network. In addition, these brightest regions are particularly flat and rounded at the edges. Si clusters properties have been intensively investigated in the past, while no reports can be found on the presence and the origin of such brilliant areas in Si/ graphite systems. A similar observation has been shown to occur in the case of silicon deposited on a $MoS_2$ substrate kept at room temperature.[20] The authors interpreted their STM measurements in terms of Si intercalation under the first $MoS_2$ layers forming a flat, bi-dimensional silicon cluster.

However, there are a number of works showing that Si intercalation can occur only if the substrate is kept or annealed at temperatures higher than 800-900K.[21-24] On the other hand, some theoretical calculations predict the formation of silicene nanosheets sandwiched between two graphene layers occurring at room temperature.[25,26] Therefore, to shed light on the nature of such flat protrusions on HOPG surface, we performed Raman measurements.

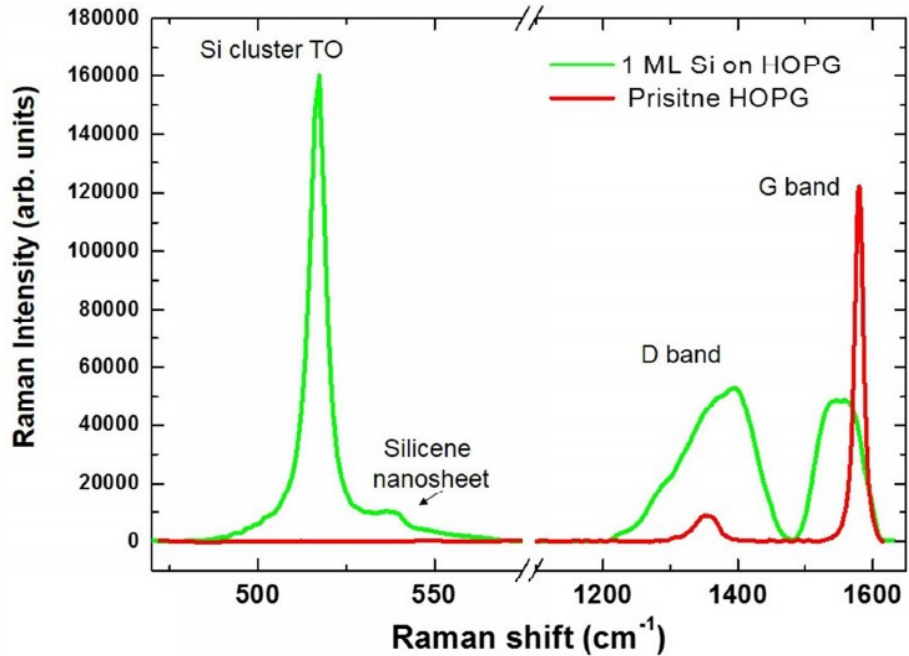

Figure 3: Raman spectra of the clean HOPG surface (red curve) and of the HOPG surface after the 1 ML Si deposition at room temperature (green curve). Note the presence of the TO resonance typical of sp$^3$ Si three-dimensional arrangement located at 517 cm$^{-1}$ and of a new peak located at 538 cm$^{-1}$ which we ascribe to silicene intercalated under the first layer of graphene. The G and D band of the HOPG are completely different from the ones of the clean HOPG. These differences are discussed in the text in terms of strain experienced by the outermost graphene layer after silicene nanosheet intercalation and presence of graphite edges.

In Fig. 3, we report the Raman spectra recorded before and after Si deposition (red and green curves, respectively). The Raman spectrum of freshly cleaved HOPG curve is constituted by the D and G Raman modes, typical of a defective area, probably related to the edge of the HOPG terraces.[27,28] The Raman spectrum after Si deposition is characterized by several resonances and the predominant feature, localized at 517 cm$^{-1}$, originates from the transverse optical phonon of sp$^3$ hybridized Si nanoclusters.[10] Close to this resonance, there is a small feature located at about 538 cm$^{-1}$ (see Fig. 3 and Fig. S2 in the SI). This new Raman resonance is located very close to that reported by Castrucci et al. peaked at 542.5 cm$^{-1}$ and ascribed to silicene formation on the HOPG surface.[11] This suggests that, also in this case, Raman spectroscopy probes the typical atomic vibrations of a silicene layer. Interestingly, no Raman resonance has been detected around 800 cm$^{-1}$, where silicon carbide modes are expected.[29] This indicates that no Si-C interaction occurred upon

Si deposition at room temperature on HOPG substrate. In addition, opposite what shown by Castrucci et al.,[11] here a dramatic modification of the graphite Raman modes is present. The G and the D bands widen as if they were composed of several contributions differently from those of clean HOPG. In particular, the G band presents components extending toward lower wavenumbers and the D band intensity dominates over the G band one. In order to clarify the effect of the silicon intercalation, the Raman spectra are fitted with Voigt curves, and the obtained Raman peak parameters are reported in Table S1 and Table S2 of the SI. The Voigt analysis reveals the doubling of the D and G peaks, showing an additional peak for each mode. In Table S2 of the SI also the parameters of the Raman modes of the sp$^3$ silicon and of silicene are reported. The D band fitting analysis shows the presence of two components (reported in the SI) due to the contribution from the edges of graphite (high amount of defects).[28] For the G band, it is well known that it is extremely sensitive to strain in carbon nanotube and graphene, owing to the phonon deformation caused by the change in lattice constant.[27]

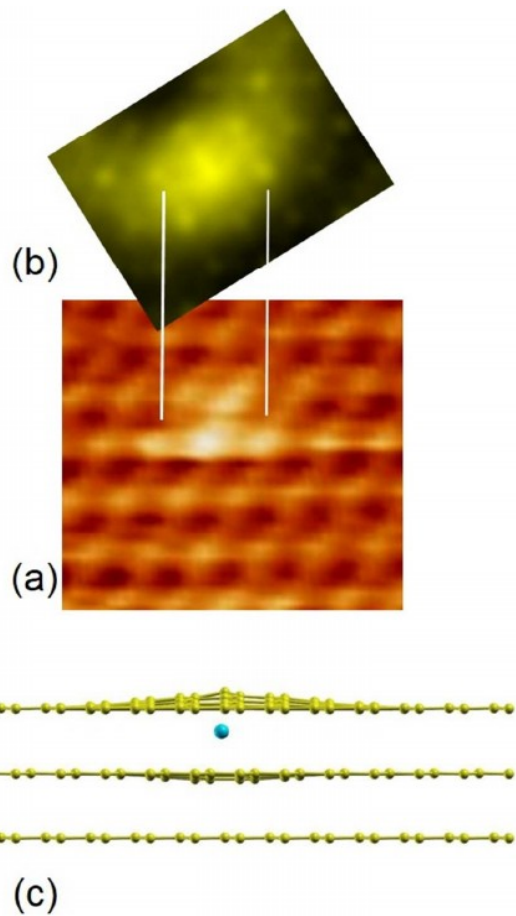

Figure 4: (a) *Ab-initio* theoretical calculation of the current image of one Si atom located underneath the outermost graphene layer of HOPG substrate. (b) Experimental STM image. (c) Sketch of the atomic vertical arrangement of the HOPG and Si atom profile as result for the calculations. Note that the presence of the Si atom produces a bump in the upper graphene layer and a depression in the one underneath, involving not only the carbon atom located on top or under the Si atom but also a number of surrounding carbon atoms.

In order to understand the origin of such STM and Raman observations, we performed *ab-initio* calculations to give a coherent interpretation of the experimental results. Fig. 4b shows the computed STM image, calculated within the Tersoff-Hamann approximation[30] due to the insertion of one Si atom under the first layer graphite slab. The theoretical current image is in good agreement with the experimental STM image of Fig. 4a. It is worth noting that the presence of the carbon atoms. Si atom produces a bump in the upper graphene layer and a depression in the one underneath, involving not only the carbon atom located on top or under the Si atom but also a number of surrounding carbon atoms, as it is seen from Fig. 4c. This effect is even more evident when we increase the number of Si atoms under the first graphene layer as shown in Fig. 5b and evidenced by the calculated line profile reported in Fig. 5c. We consider here a system consisting of

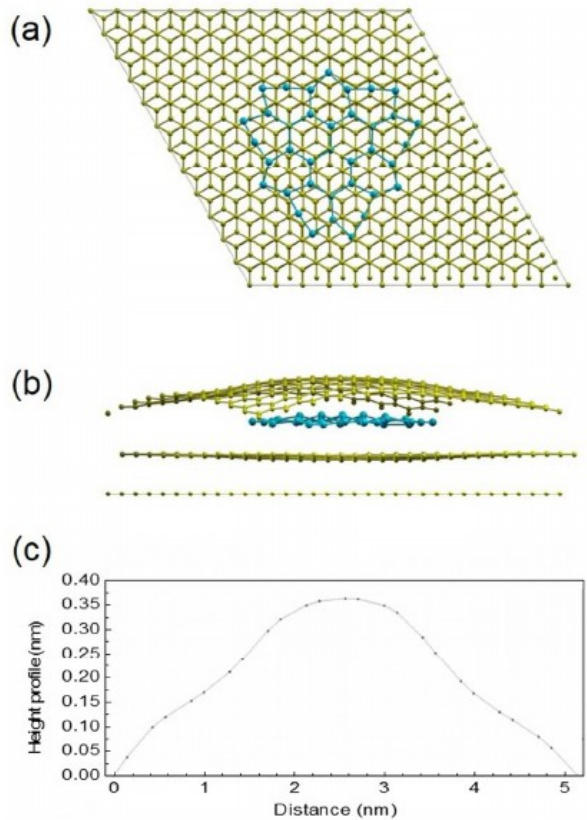

Figure 5: Sketch of the planar (a) and vertical (b) arrangement of the Si and C atoms as results from the *ab-initio* calculations for a system consisting of a 12×12×3 C-slab with 37 Si-atoms inserted under the first C-layer. (c) Theoretical height profile of the outermost graphene layer showing the effect induced by the silicene nanosheet. The height of the outermost graphene layer over the silicene nanostructure is about 0.15 nm quite constant over all the Si atom layer.

a 12×12×3 C-slab with 37 Si-atoms inserted under the first C-layer. Although this system is very large and already contains 901 atoms, the top layer was detached from the slab during the geometry optimization. This means that a larger number of in-plane unit cells is required for such a number of Si-atoms. However, instead of increasing the size of C-slab, we fixed one C-atom of the first layer (at the corner of unit cell) to avoid the first layer detachment during the introduction of Si atoms. The 37 Si atoms were initially positioned as for ideal free-standing silicene structure (i.e. honeycomb network) with a buckling between nearest neighbor atoms of 0.042 nm.[7,31] The resulting relaxed Si nanostructure has a typical silicene arrangement and remains located under the first graphene layer. The corresponding theoretical height profile (see Fig. 5c) of the outermost graphene layer over the silicene nanostructure is about 0.15 nm quite constant over all the Si atom layer (see line profile in Fig. 5c). This is in very good agreement with the experimental results reported in

Fig. 2. In addition, charge density calculation shows that no bonds are present between Si and C atoms (see Fig. 6). These calculations give hints of silicene nanostructure formation under the first graphene layer.

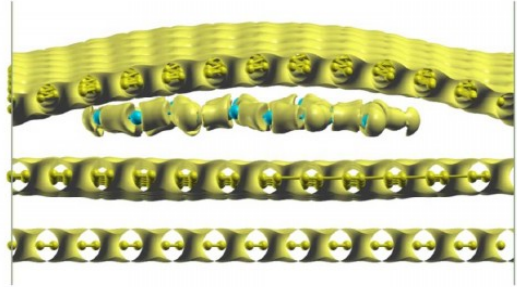

Figure 6: Charge density calculation for a system consisting of a 12×12×3 C-slab with 37 Si-atoms inserted under the first C-layer. It shows that no bonds are present between Si and C atoms.

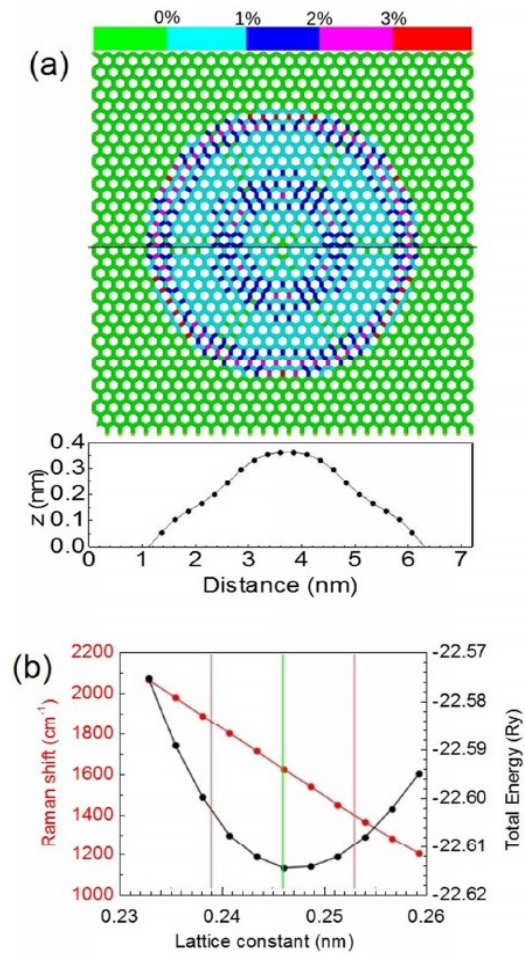

Figure 7: (a) Distribution of the calculated tensile strain of the C atoms on top of the silicene nanosheet. Different colors represent the percentage of strain, which amounts to values between the 0 to 4 % of the C-C bond length of free-standing graphene. (b) Calculated graphite G band Raman shift as a function of the lattice parameter. The calculations were performed using density functional perturbation theory[29] with the general set of parameters, used through the article, apart the k-point grid was selected to be 25×25×1. The green line indicates the energy stable unstrained position of the graphite lattice parameter, giving rise to a Raman shift equal to 1626 cm$^{-1}$.

Fig. 7a reports the distribution of the calculated tensile strain of the C atoms on top of the silicene nanosheet. Different colors represent the percentage of strain, which varies between the 0 to 4 % of the C-C bond length of freestanding graphene. It is worth noting that the highest amount of strain is located close to the edge of the silicene nanosheet and at the most external border of the resulting graphene bubble. Due to the low value of the strain, it is very hard to experimentally measure variation in C atom nearest neighbor distance with STM. However, these strain values have a dramatic effect on the vibration of the atoms and therefore in the corresponding Raman features. In Fig. 7b, we report the calculated graphite G band Raman shift as a function of the lattice parameter. The calculations were performed using density functional perturbation theory [32] with the general set of parameters, used through the article, apart the k-point grid was selected to be 25×25×1. The green line indicates the energy stable unstrained position of the graphite lattice parameter, giving rise to a Raman shift equal to 1626 cm$^{-1}$. Within our model and parameters selected, this value appears to be greater than that observed experimentally. We performed a number of test calculations with higher wave-function cutoff and denser k-point grids, which showed us a better approach to experimental value of G line position, but had no effect on dependence of G-band position on lattice constant. Therefore, we decided to proceed with the general parameter set to be coherent with the rest of results. The calculations show that tensile strains, traducing in longer lattice parameter than the equilibrium one, shift the Raman feature of the G band towards lower wavenumbers. The major part of carbon atoms have a strain percentage between the 1 and 2 % which corresponds to Raman downshift of about 50 and 100 cm$^{-1}$, respectively. This calculation allows to account for the huge FWHM of about 100 cm$^{-1}$ of the experimental G band measured for the spectrum showing also the Raman feature located around 538 cm$^{-1}$.

The joint theoretical and experimental study reported gives new insight into the formation of silicene on HOPG. STM and Raman results show that at room temperature, Si atoms intercalate under the top graphene layer of the HOPG substrate. This phenomenon involves many Si atoms that nucleate in the form of silicene nanosheets. These sandwiched silicene 2D nanosheets do not interact with the top and bottom C layers. The Raman feature located at 538 cm$^{-1}$ suggests the formation of silicene nanosheets. In correspondence, the upper graphene layer experiences a huge tensile strain which can reach 4% producing dramatic changes in the G band as measured in the Raman spectrum. This is another fundamental piece of information in the highly debated and still open question of the synthesis, existence and stability of silicene on inert substrates.

**Acknowledgments**

I. K., O.P., M.D.C., P.C., M.S. and M.S. would like to acknowledge the European Community for the HORIZON2020 RISE Project CoExAN (GA644076). CPU time was granted by CINECA HPC center.

Supplementary Information

# Scanning tunneling microscopy and Raman evidences of silicene nanosheets intercalated into graphite surface at room temperature

I.Kupchak [1], F.Fabbri[2], M.De Crescenzi[3], M. Scarselli [3], M. Salvato[3], T.Delise [3], I.Berbezier [4], O.Pulci [3] and P. Castrucci [3,*]

[1] *V. E. Lashkaryov Institute of Semiconductor Physics, National Academy of Sciences of Ukraine, Ukraine*

[2] *Center for Nanotechnology Innovation c/o NEST, Istituto Italiano di Tecnologia, Pisa 56127, Italy*

[3] *Dipartimento di Fisica, Università di Roma "Tor Vergata", 00133 Roma, Italy.*

[4] *CNRS, Aix-Marseille Université, IM2NP, UMR 7334, Campus de St. Jérome, 13397 Marseille, France.*

[*] e-mail: paola.castrucci@roma2.infn.it

Abstract:

Graphene is expected to be an excellent substrate to grow two-dimensional silicene layer, for its inertness and its structural honeycomb network which serves as template. Si atoms are not predicted to swap with C atoms at room temperature and form intercalated Si clusters. In this work, we demonstrated by using scanning tunneling microscopy (STM) and Raman spectroscopy that for a highly oriented pyrolitic graphite (HOPG) surface, Si atoms deposited at room temperature arrange in two configurations: as silicene nanosheets under the first atomic layer of HOPG and as three dimensional clusters upon HOPG surface. The Raman spectra reveal a structure located at 538 cm$^{-1}$ which we ascribe to the presence of sp$^2$ Si hybridization. In addition, the silicon deposition induces several modification to graphite D and G Raman modes, interpreted as an experimental evidence of the intercalation of the silicene nanosheets. The silicon intercalation at room temperature occurs at the HOPG terrace and it detaches only the outermost graphene layer producing a graphene strong tensile strain mainly concentrated around silicene nanosheet, as evidenced by Raman spectroscopy. Theoretical calculations on the structure, stability and Raman resonances of the silicene nanosheet sandwiched among graphene and graphite layers have been done to support the experimental STM and Raman observations.

In Fig. S1 we report the STM image of an extended area of the HOPG surface after the deposition of 1 ML of Si at room temperature. Note that there is a coexistence of three-dimensional Si clusters and two-dimensional nanosheets, which at more resolved STM measurements show the hexagonal network and the lattice parameter typical of graphene. Two height profiles are reported in the right panel of the figure, exhibiting the height modulation for the sp3 Si cluster (lower curve) and the two-dimensional bubble (upper curve, yellow line on the image).

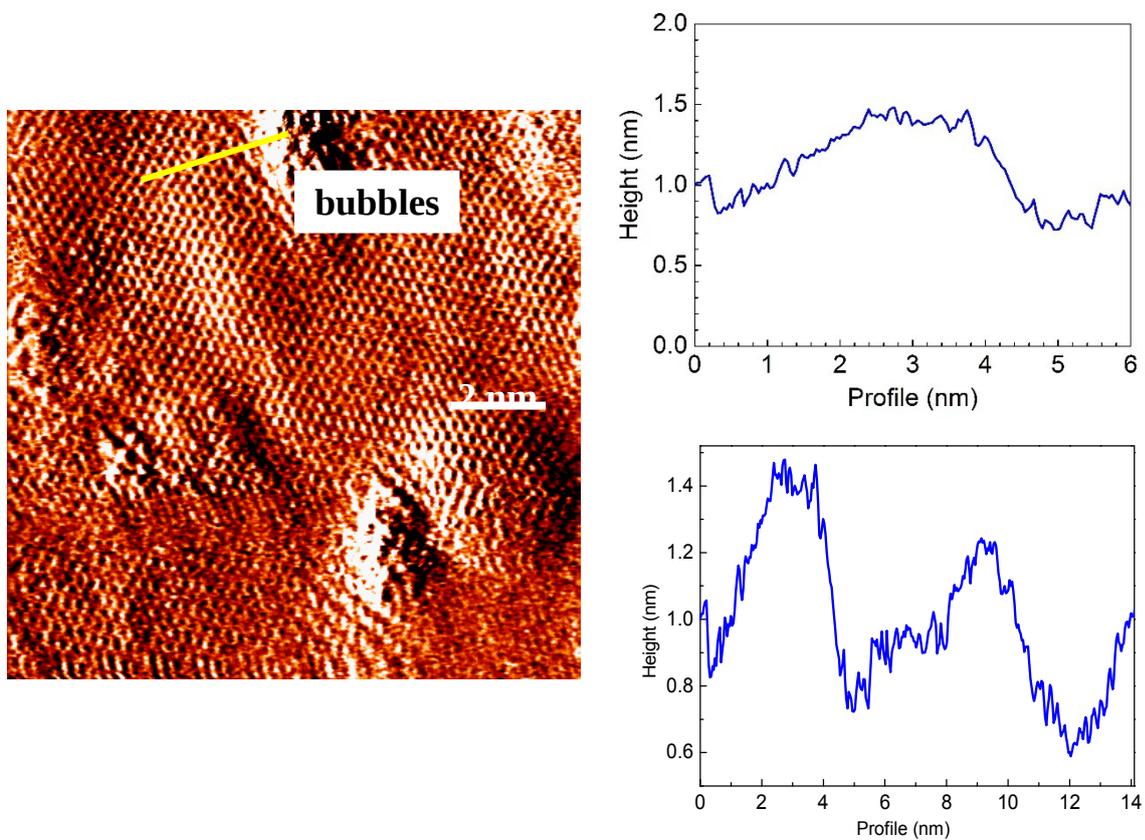

Fig. S1: Left panel: STM image ($V_{bias}$ = 0.3V, $I_{tunn}$ = 0.4 nA) of an extended area of the HOPG surface after the deposition of 1 ML of Si at room temperature. Right panel: height profiles of the Si sp3 cluster (lower curve) and of the two-dimensional bubble (upper curve, yellow line on the STM image).

Fig. S2: Raman spectra deconvolution with Voigt curves. Raman peak parameters are reported in

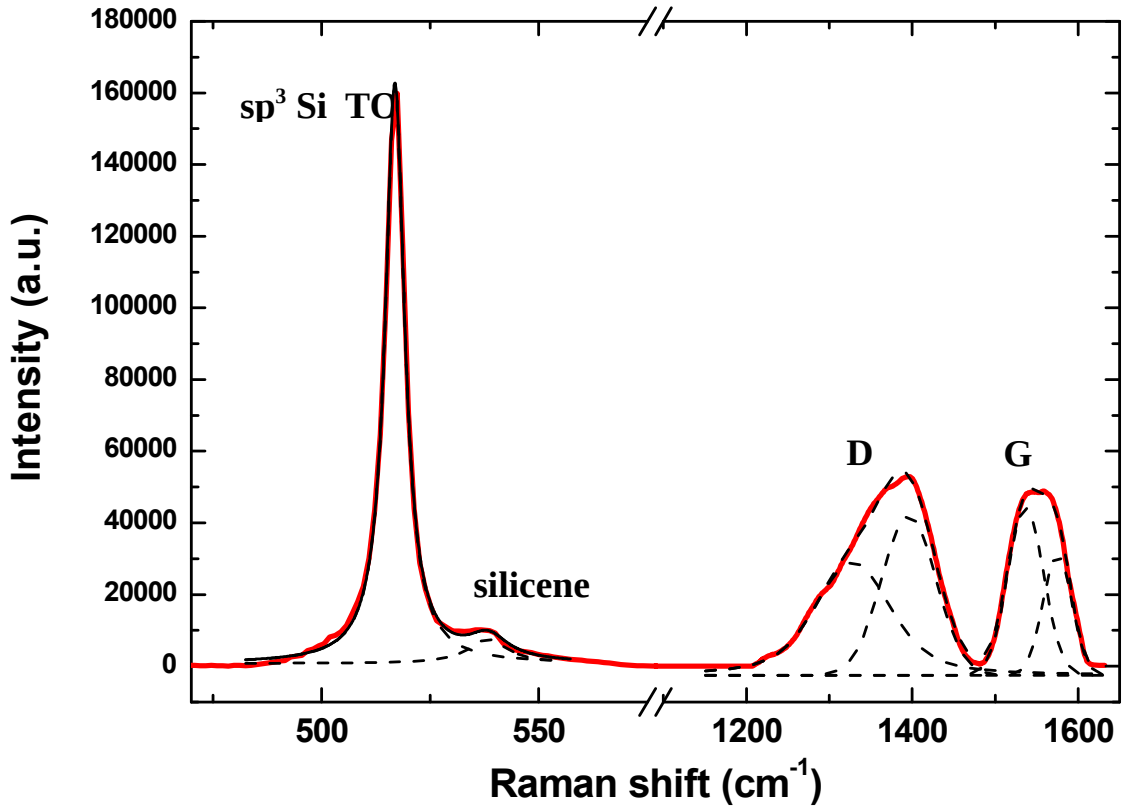

Table S1 and Table S2. The Voigt deconvolution reveals the doubling of the D and G peaks, showing an additional peak for each mode. In Table S2 presents the parameters of the Raman modes of the sp3 silicon (517 cm$^{-1}$) and of silicene located at 538 cm$^{-1}$

Table S1

|  | **Before Silicon Deposition** | | | **After Silicon Deposition** | | |
| --- | --- | --- | --- | --- | --- | --- |
|  | Raman Shift (cm$^{-1}$) | FWHM (cm$^{-1}$) | Intensity x 10$^5$ (cps) | Raman Shift (cm$^{-1}$) | FWHM (cm$^{-1}$) | Intensity x 10$^5$ (cps) |
| G mode | 1579 | 15 | 25.6 | 1575 | 41 | 14.8 |
|  |  |  |  | 1536 | 51 | 25.3 |
| D mode | 1359 | 43 | 6.2 | 1328 | 108 | 43.9 |
|  |  |  |  | 1395 | 77 | 36.4 |

Table S2

|  | Raman Shift (cm$^{-1}$) | FWHM (cm$^{-1}$) | Intensity x 10$^5$ (cps) |
| --- | --- | --- | --- |
| Si TO | 517 | 5.4 | 13.7 |
| Silicene | 538 | 11.2 | 1.2 |